\documentclass[sn-mathphys,Numbered]{sn-jnl}

\usepackage{graphicx}%
\usepackage{multirow}%
\usepackage{amsmath,amssymb,amsfonts}%
\usepackage{amsthm}%
\usepackage{mathrsfs}%
\usepackage[title]{appendix}%
\usepackage{xcolor}%
\usepackage{textcomp}%
\usepackage{manyfoot}%
\usepackage{booktabs}%
\usepackage{algorithm}%
\usepackage{algorithmicx}%
\usepackage{algpseudocode}%
\usepackage{listings}%
\usepackage{lineno}%

\begin{document}

\title[Highly efficient channeling of single photons into guided modes of optical nanocapillary fibers]{Highly efficient channeling of single photons into guided modes of optical nanocapillary fibers}

\author[1]{\fnm{Bashaiah} \sur{Elaganuru}}\email{18phph19@uohyd.ac.in}

\author[1]{\fnm{Resmi} \sur{M}}\email{19phph18@uohyd.ac.in}

\author*[1]{\fnm{Ramachandrarao} \sur{Yalla}}\email{rrysp@uohyd.ac.in}

\affil[1]{\orgdiv{School of Physics}, \orgname{University of Hyderabad}, \orgaddress{\city{Hyderabad}, \postcode{500046}, \state{Telangana}, \country{India}}}

\abstract{We report numerically the efficient channeling of single photons from a single quantum emitter into guided modes of optical nanocapillary fibers (NCFs). The NCF is formed of a liquid core optical nanofiber with inner and outer diameters. We optimize the inner and outer diameters of the NCF filled with water medium by placing a single dipole source (SDS) inside. The maximum channeling efficiency of $52$\% is found when the radially polarized SDS is placed at the center of the NCF filled with the water medium. The optimum inner and outer diameters of the NCF are $100$ nm and $360$ nm for the emission wavelength of  $620$ nm, respectively. Additionally, we investigate the SDS position dependence inside the NCF considering experimental ambiguity in placing a single quantum emitter inside the NCF. We found that the channeling efficiency remains almost constant for the water medium at the optimum condition. The present platform may open a novel route for generating single photons in quantum technologies and detecting single cells in bio-sensing.}


\maketitle

\section{Introduction}\label{sec1}
 In quantum information processing and communication, single photons are employed as information carriers as they are the ideal choice. A quantum network consists of quantum nodes for storing single photons and quantum channels for single photon transmission. Single-mode optical fibers can serve as quantum channels and single quantum emitters as quantum nodes in a physical implementation of a quantum network. Therefore, efficient channeling of single photons into a single-mode optical fiber is necessary. The generation of single photons nowadays is achieved using various protocols based on single atoms \cite{hood2000atom}, single ions \cite{barros2009deterministic}, single molecules \cite{fleury2000nonclassical}, single quantum dots \cite{michler2000quantum}, single vacancy centers in nano-diamonds \cite{kurtsiefer2000stable}, single defect in silicon carbide \cite{aharonovich2016solid}, rare-earth-ion impurities in yttrium aluminum garnet/yttrium orthosilicate \cite{aharonovich2016solid}, and two-dimensional materials \cite{aharonovich2016solid}. 

However, the production of single photons is useless without an efficient manipulation and control manner. Efficiently channeling single photons from a single quantum emitter into a particular mode is a challenging task \cite{kimble2008quantum,bremer2022fiber}. Various protocols have been proposed and developed for efficiently collecting single photons. Examples include micropillar cavities \cite{solomon2001single}, solid immersion lens \cite{schroder2011ultrabright}, photonic crystal cavities \cite{shambat2011optical}, diamond waveguides \cite{patel2016efficient}, plasmonic metal nanowires \cite{akimov2007generation}, optical nanofibers (ONFs)   \cite{nayak2018nanofiber,vetsch2010optical,yalla2012efficient,fujiwara2011highly,
le2005spontaneous,klimov2004spontaneous,chonan2014efficient,yonezu2017efficient,morrissey2013spectroscopy,yang2023generating,yalla2022integration}, and optical nanofiber tips (ONFTs)\cite{resmi2023efficient,chonan2014efficient,das2023efficient}. Other than ONF/ONFT, the subsequent coupling of fluorescence photons into a single-mode fiber reduces the actual collection efficiency due to the mode mismatch. Regarding atoms and ions, it is possible to efficiently control the production of single photons on demand. However, it requires complicated experimental setups and intermittent loading of atoms or ions. Regarding solid-state quantum emitters, 
spectral diffusion and other broadening result in photon distinguishability from the same quantum emitter/various emitters. Collecting single photons from quantum emitters in the high refractive index host materials is challenging. \\

In addition to efficient manipulation, efficient transmission lines between quantum nodes are also essential for applications in quantum technologies. In such an application, efficient channeling of single photons into single-mode optical fibers (SMFs), playing the role of transmission lines, is required. To achieve this, nanostructured systems have been proposed and demonstrated, including sub-wavelength diameter silica fibers termed ONFs and ONFTs \cite{vetsch2010optical,yalla2012efficient,fujiwara2011highly,
le2005spontaneous,klimov2004spontaneous,chonan2014efficient,yonezu2017efficient}. Moreover, for application in quantum network \cite{kimble2008quantum}, in-line method i.e. fiber integrated light-matter interfaces such as those realized by ONFs/ONFTs are advantageous since automatic channeling to the SMF was achieved \cite{le2005spontaneous,klimov2004spontaneous,fujiwara2011highly,yalla2012efficient}. However, in these efforts, a maximum channeling efficiency ($\eta$) was achieved up to $28$\% for the radially polarized single dipole source (SDS) \cite{le2005spontaneous,klimov2004spontaneous}. The maximum $\eta$-value up to $20$\% has been experimentally demonstrated for single photons from a single quantum dot directly into the SMF considering randomly oriented dipole \cite{yalla2012efficient}. Note that the single quantum dot was placed on the surface of the ONF. It has been numerically shown that a maximum $\eta$-value of 38\% from the radially polarized SDS to the SMF was achieved using the ONFT \cite{resmi2023efficient,chonan2014efficient}. Note that the SDS was placed on the facet of the tip. The feasibility of the simulation results has been experimentally shown by placing quantum dots on the facet of the ONFT \cite{m2024channeling}. Note that the interaction strength has been limited due to the positioning of the SDS on the surface of the ONF/ONFT. \\ 

One can enhance the $\eta$-value in two ways: one is creating a cavity on the ONF and the other is placing the single quantum emitter inside the ONF. Regarding the cavity on the ONF, the maximum $\eta$-value of $65$\% has been experimentally demonstrated using a composite photonic crystal cavity on the ONF \cite{yalla2014cavity}. Note that a single quantum dot was placed on the surface of the ONF only, which limits the interaction. Regarding placing the single quantum emitter inside, a new type of hollow core fiber with sub-micron diameter termed as capillary fibers has been proposed and experimentally demonstrated \cite{faez2014coherent}. However, these efforts demonstrated the $\eta$-value up to $18$\% only. This is due to the thicker outer diameter of the capillary fiber, leading to weak confinement of the field inside the capillary fiber. The systematic study of $\eta$-value for the hollow/liquid core ONF with inner and outer diameters termed optical nanocapillary fibers (NCFs) has not been reported yet. Note that the inner and outer diameters of the NCF are in sub-wavelength. The position dependence of the SDS inside the NCF, keeping in view of experimental ambiguity, has not been reported yet.\\

In this paper, we use numerical simulations to report the efficient channeling of single photons from a single quantum emitter into guided modes of the optical NCF. We perform numerical simulations to optimize the inner and outer diameters of the NCF filled with liquid medium by placing the SDS inside, which was not possible in the case of the ONF. A maximum  $\eta$-value of  $52$\% is found when the radially polarized SDS is placed at the center of the NCF filled with water medium. For the emission wavelength of  $620$ nm, the optimum inner and outer diameters of the NCF filled with water are $100$ nm and $360$ nm, respectively.  Additionally, we investigate the position dependence of the SDS inside the NCF keeping in view of experimental ambiguity. Simulated results show that the $\eta$-value remains almost the same for the water medium while changing the position of the SDS inside the NCF.\\

\section{Methodology}\label{sec2}
\begin{figure}[h]%
	\centering
	\includegraphics[width=0.8\textwidth]{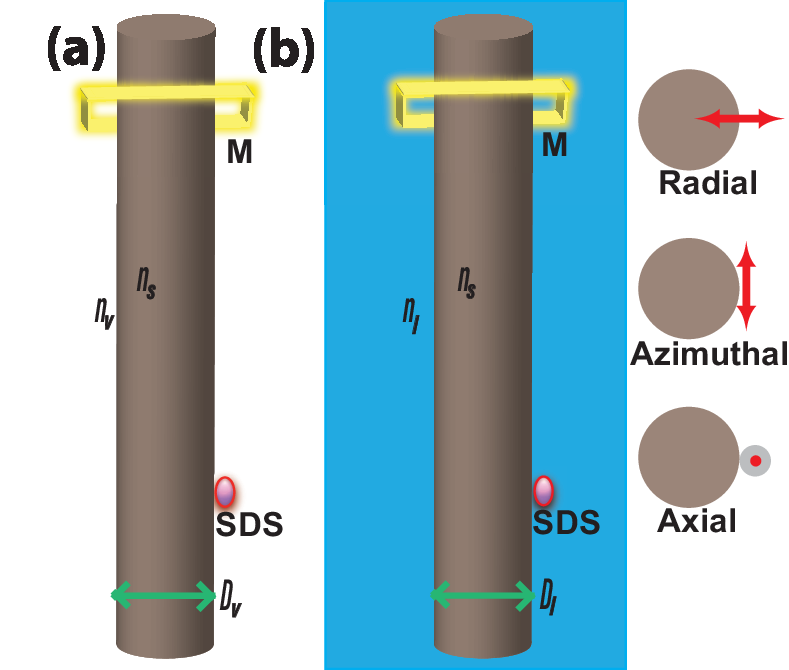}
	\caption{(a) and (b) show the conceptual sketches of the ideas for the optical nanofiber (ONF) placed in vacuum and liquid medium, respectively. $D_v$ and $D_l$ are the ONF diameters when the surrounding medium is vacuum and liquid, respectively. $n_s$, $n_v$, and $n_l$ are refractive indices of silica, vacuum, and liquid, respectively. SDS and M denote the single dipole source and monitor, respectively. The SDS is placed on the surface of the ONF. The insets show the radial, azimuthal, and axial polarizations corresponding to the SDS directed perpendicular, tangent, and parallel to the cylindrical surface, respectively.}\label{fig1}
\end{figure}

We perform numerical simulations using Ansys lumerical finite difference time domain (Ansys, FDTD Package) platform \cite{taflove2005computational, schneider2010understanding}.
First, we perform simulations to determine the $\eta$-value into guided modes of the ONF in vacuum and liquid medium for reference. Conceptual sketches of the ideas are shown in Figs. \ref{fig1} (a) and (b) for vacuum and liquid medium, respectively. We determine the $\eta$-value for different diameters of the ONF ($D_v$/$D_l$) in both mediums. The SDS is placed on the surface of the ONF. We set the polarization of the SDS to be radial/azimuthal/axial. We place the power monitor (M) sufficiently far from the SDS position to determine the coupled power ($P_c$). $P$ and $P_0$ are the total power emitted by the SDS in the presence of the ONF and in the vacuum environment, respectively. Thus, the channeling efficiency is defined as $\eta$= $P_c/P$. We determine the $\eta$-value for different polarizations while sweeping the $D_v$/$D_l$-values  of the ONF. For each case, we find the optimum $D_v$/$D_l$-values by maximizing the $\eta$-value into the guided mode of the ONF.

\begin{figure}[h]%
	\centering
	\includegraphics[width=0.8\textwidth]{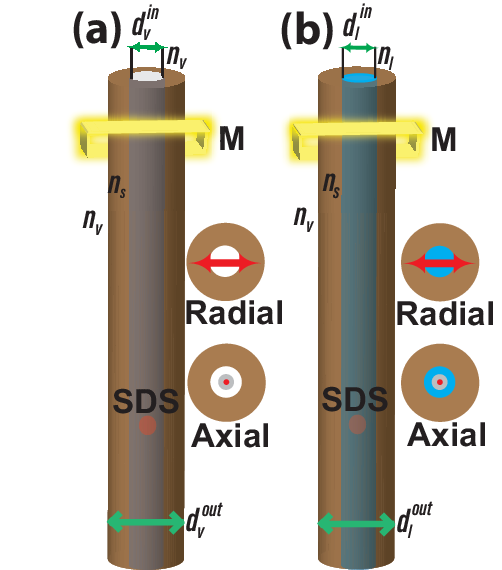}
	\caption{(a) and (b) show the conceptual sketches of the ideas for the nanocapillary fiber (NCF) filled with vacuum and liquid medium, respectively. $d^{in}_v$/$d^{in}_l$ and $d^{out}_v$/$d^{out}_l$ are the inner and outer diameters of the NCF filled with vacuum/liquid, respectively.  $n_s$, $n_v$, and $n_l$ are refractive indices of silica, vacuum, and liquid, respectively. SDS and M denote the single dipole source and monitor, respectively. The SDS is placed inside the NCF. The insets show radial and axial polarizations corresponding to the SDS directed perpendicular and parallel to the cylinder axis, respectively.}\label{fig2}
\end{figure}

Next, we perform simulations to determine the $\eta$-value for the NCF based on the optimum ONF diameters ($D_v$/$D_l$). Conceptual sketches of the NCF-filled with vacuum and liquid medium are shown in Figs. \ref{fig2} (a) and (b), respectively. The SDS is placed inside the NCF. We determine the  $\eta$-value for different polarizations while sweeping inner and outer diameters ($d^{in}$ and $d^{out}$) of the NCF in both mediums. The vacuum and liquid-filled NCFs are defined as hollow-core NCF and liquid-core NCF, respectively. For each case, we find the optimum $d^{in}$ and $d^{out}$-values by maximizing the $\eta$-value into the guided mode of the NCF.

ONF, NCF, SDS, and M are placed inside the three-dimensional simulation area of 6$\times$6$\times$35 $\mu m^3$ enclosed in the perfectly matched layers (PMLs) to avoid reflections. The length of the ONF/NCF is set at $45$ $\mu m$ so that they can be treated as infinitely long. The maximum ONF/NCF diameter is chosen within the simulation region. The SDS is placed 5 $\mu m$ away from the PML. The M is placed 25 $\mu m$ away from the SDS and 5 $\mu m$ away from the PML. In the case of the ONF, the radial, azimuthal, and axial polarizations corresponding to the SDS are directed perpendicular, tangent, and parallel to the cylindrical surface, respectively as clearly shown in the inset of Fig. \ref{fig1} (b). In the case of the NCF, the radial and axial polarizations corresponding to the SDS are directed perpendicular and parallel to the cylinder axis, respectively as clearly shown in the insets of Fig. \ref{fig2}. We set the emission wavelength of the SDS at 620 nm, corresponding to the emission wavelength of the quantum emitter \cite{shafi2018hybrid}. In the case of the ONF, refractive indices of a cylinder and surrounding mediums are set for silica ($n_s$) and vacuum ($n_v$)/liquid ($n_l$), respectively. In the case of the hollow core NCF, refractive indices of the outside of the cylinder, cylinder, and inside of the cylinder are set for vacuum ($n_v$), silica ($n_s$), and vacuum ($n_v$), respectively. In the case of the liquid core NCF, refractive indices of the outside of the cylinder, cylinder, and inside of the cylinder are set for vacuum ($n_v$), silica ($n_s$), and liquid ($n_l$), respectively.

\section{Results}\label{sec3}

\begin{figure}[h]%
	\centering
	\includegraphics[width=\textwidth]{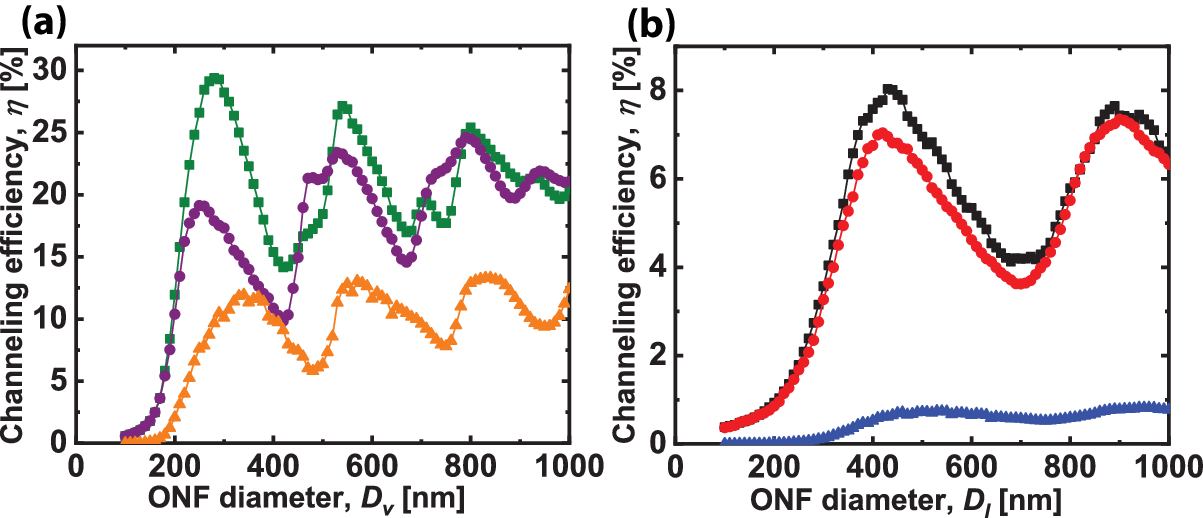}
	\caption{(a) and (b) show summaries of $\eta$-values as a function of the optical nanofiber (ONF) diameters ($D_v$/$D_l$) for three polarizations when the ONF is placed in vacuum and water, respectively. Green/black squares, purple/red circles, and orange/blue triangles correspond to the radial, azimuthal, and axial polarizations, respectively.}\label{fig3}
\end{figure}

Figure \ref{fig3} (a) shows the dependence of $\eta$-value as a function of the ONF diameter ($D_v$) for the SDS placed on the surface of the ONF. Horizontal and vertical axes correspond to the $D_v$-value and $\eta$-value, respectively. The SDS along the radial, azimuthal, and axial polarizations are shown by green squares, purple circles, and orange triangles, respectively. One can readily see the maximum $\eta$-value is for the radial polarization. This is due to the strong confinement of the field in the radial direction. For the radial, azimuthal, and axial polarizations, we found the maximum $\eta$-value of 29\%, 18\%, and 9\%, respectively. The maximum $\eta$-value of 29\% occurred at the $D_v$-value of 280 nm, corresponding to the fiber size parameter of $1.42$. 

Figure \ref{fig3} (b) shows the dependence of $\eta$-value as a function of the ONF diameter 
($D_l$) for the SDS placed on the surface of the ONF and the surrounding medium is water. Horizontal and vertical axes correspond to $D_l$-value and $\eta$-value, respectively. The SDS along the radial, azimuthal, and axial polarizations are shown by black squares, red circles, and blue triangles, respectively. One can see that the maximum $\eta$-value occurred for the radial polarization. For the radial, azimuthal, and axial polarizations, we found the maximum $\eta$-value of 8\%, 7\%, and 1\%, respectively. The maximum $\eta$-value of 8\% occurred at the $D_l$-value of 430 nm, corresponding to the fiber size parameter of $2.17$.

\begin{figure}[h]%
	\centering
	\includegraphics[width=\textwidth]{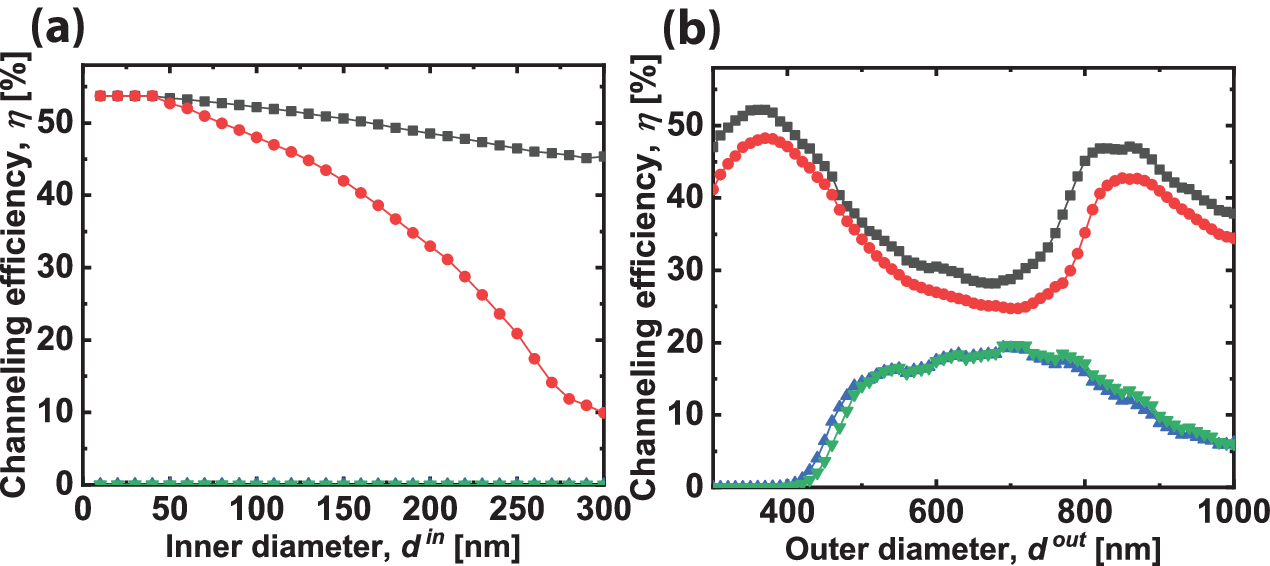}
	\caption{(a) The summary of $\eta$-values as a function of the inner diameter ($d^{in}$) of the nanocapillary fiber (NCF) filled with vacuum and water for two polarizations. The outer diameter ($d^{out}$) of the NCF is fixed at 360 nm. (b) The summary of $\eta$-values as a function of $d^{out}$-values of the NCF filled with vacuum and water for two polarizations. The $d^{in}$-value of NCF is fixed at 100 nm. Black squares and red circles denote the radial polarization in water and vacuum, respectively. Blue and green triangles denote the axial polarization in water and vacuum, respectively.}\label{fig4}
\end{figure}
We perform simulations to determine the $\eta$-value for the NCF outer diameter based on the optimum $\eta$-value of ONF diameters in free space and water medium. The NCF outer diameter is chosen as the average value of optimum diameters of ONF in both free space and water medium. Then, we sweep the inner diameters while the outer diameter is fixed at the average value. Then, we fix the inner diameter of the NCF at the value where there is no significant decay in the $\eta$-value and consider the experimental realization of such a thin capillary hole. Then, we sweep the outer diameter of the NCF and find the optimum outer diameter of the NCF.

Figure \ref{fig4} (a) shows the dependence of $\eta$-value as a function of the NCF inner diameters ($d^{in}$) filled with vacuum/water. Note that the SDS is placed at the center of the NCF. The outer diameter ($d^{out}$) of the NCF is fixed at $360$ nm. Based on the previous simulation results, The $d^{out}$-value is chosen as the average value of optimum $D_v$ and $D_l$-values. The $d^{in}$-value of the NCF varies from $10$ nm to $300$ nm. The vertical axis corresponds to the $\eta$-value. Black squares and red circles represent the radial polarization of the SDS in water and vacuum, respectively. Blue and green triangles represent the axial polarization of the SDS in water and vacuum, respectively. In both cases, one can readily see that the maximum value occurred for the radial polarization. For the radial polarization, one can readily see that the maximum value occurred for the water medium. We found the maximum $\eta$-value of 53\%. Note that the $\eta$-value for the radial and azimuthal polarizations are the same when the SDS is at the center of the NCF.

Figure \ref{fig4} (b) shows the dependence of $\eta$-value as a function of the NCF outer diameter ($d^{out}$). The $d^{in}$-value of the NCF is fixed at $100$ nm considering the experimental feasibility. The $d^{out}$-value of the NCF is varied from $300$ nm to $1000$ nm. The vertical axis corresponds to the $\eta$-value. Black squares and red circles represent the radial polarization in water and vacuum, respectively. Blue and green triangles represent the axial polarization in water and vacuum, respectively. One can readily see the maximum $\eta$-value occurred for the radial polarization. We found the maximum $\eta$-value of 52\% at $d^{out}$-value of $360$ nm corresponding to the fiber size parameter of $1.82$. In the case of vacuum, the maximum $\eta$-value of 48\% was found at $d^{out}$-value of $370$ nm corresponding to the fiber size parameter of $1.87$. 

\begin{figure}[h]%
	\centering
	\includegraphics[width=\textwidth]{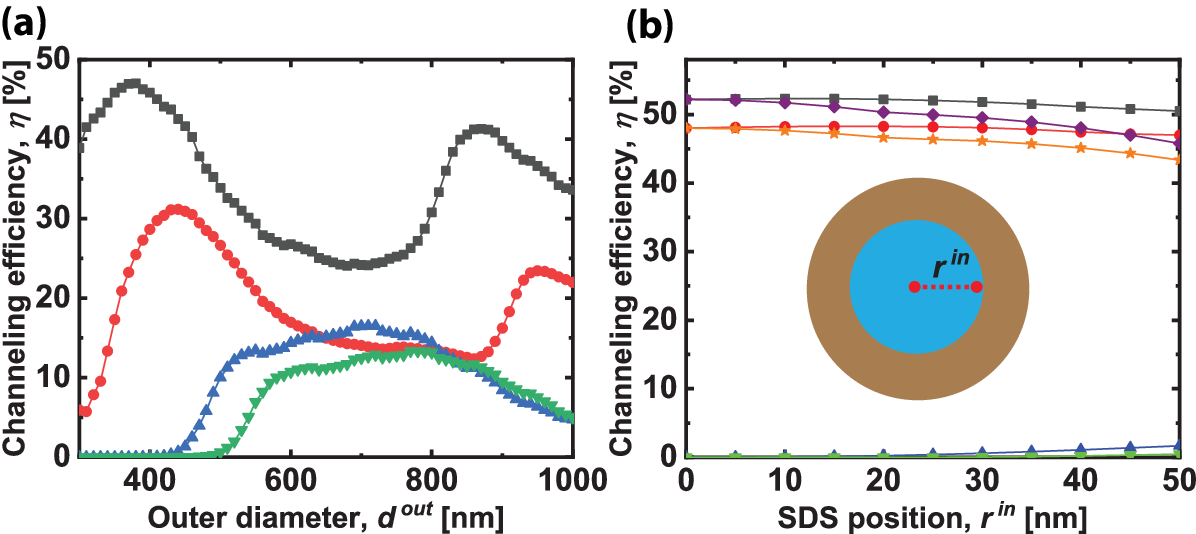}
	\caption{(a) The summary of $\eta$-values as a function of outer diameter ($d^{out}$) of the nanocapillary fiber (NCF) filled with vacuum and water for two polarizations. The inner diameter ($d^{in}$) of the NCF is fixed at 250 nm. Black squares and red circles denote the radial polarization in water and vacuum, respectively. Blue and green triangles denote the axial polarization in water and vacuum, respectively. (b) The dependence of $\eta$-value as a function of the position ($r^{in}$) of the single dipole source (SDS) for three polarizations. The inner and outer diameters of the NCF are fixed at 100 nm and 360 nm, respectively. Black squares, purple diamonds, and blue triangles are for the radial, azimuthal, and axial polarizations in the water medium, respectively. Red circles, orange stars, and green triangles are for the radial, azimuthal, and axial polarizations in the vacuum, respectively. The inset shows the schematic for the position of the SDS.}\label{fig5}
\end{figure}

Considering experimental feasibility, we investigate the choice of $d^{in}$-value.  Figure \ref{fig5} (a) shows the dependence of $\eta$-value as a function of the outer diameter ($d^{out}$) of the NCF. The $d^{in}$-value of the NCF is fixed at $250$ nm. The $d^{out}$-value of the NCF varies from $300$ nm to $1000$ nm. The vertical axis corresponds to the $\eta$-value. Black squares and red circles represent the radial polarization in water and vacuum, respectively. Blue and green triangles represent the axial polarization in water and vacuum, respectively. One can readily see the maximum $\eta$-value occurred for the radial polarization. We found the maximum $\eta$-value of 47\% at the $d^{out}$-value of $380$ nm corresponding to the fiber size parameter of $1.92$. In the case of vacuum, the maximum $\eta$-value of 31\% was found at the $d^{out}$-value of $440$ nm corresponding to the fiber size parameter of $2.23$.

Also, we investigate the effect of the position ($r^{in}$) of the SDS inside the NCF by considering the experimental ambiguity in placing a single quantum emitter. Figure \ref{fig5} (b) shows the dependence of $\eta$-value as a function of $r^{in}$-value. The $d^{in}$ and $d^{out}$-values of the NCF are kept constant at $100$ nm and $360$ nm, respectively. The horizontal axis corresponds to the $r^{in}$-value, which varies from 0 nm (center) to 50 nm (inside edge) as shown in the inset of Fig. \ref{fig5} (b). The vertical axis corresponds to the $\eta$-value. Black squares, purple diamonds, and blue triangles are for the radial, azimuthal, and axial polarizations in the water medium, respectively. Red circles, orange stars, and green triangles are for the radial, azimuthal, and axial polarizations in the vacuum, respectively. In both cases, one can readily see that the $\eta$-value corresponding to the radial polarization is maximum. Note that in both cases of the axial polarization, the $\eta$-value is close to 1\%.

\section{Discussion}\label{sec4}
The single-mode condition is defined by the parameter $V$= $ka\sqrt{n_{1}^{2} - n_{2}^{2}}$$<$ 2.405, where $a$ is the ONF radius, $k$= $2\pi$/$\lambda$, $\lambda$ is wavelength of the light, and $n_{1}$ and $n_{2}$ are refractive indices of the core and clad, respectively \cite{le2004field,nayak2018nanofiber,okamoto2021fundamentals}.
To satisfy the single-mode condition for the emission wavelength of $620$ nm, the maximum ONF diameters ($D_v$/$D_l$) are $450$ nm and $820$ nm for the clad medium of vacuum and water,  respectively. The maximum $D_v$ and $D_l$-values for single mode condition differ significantly. This is due to the significant difference in the refractive index between the silica and vacuum compared to silica and water.  As seen clearly in Figs. \ref{fig3} (a) and (b), that multimode behavior appears at the expected ONF diameters, resulting in increasing the $\eta$-value for the thicker diameter of the ONF.

As seen in Figs. \ref{fig3} (a) and (b), the optimum ONF diameters in a single-mode regime differ for three polarizations. The optimum $D_v/D_l$-values are 250/420 nm, 280/430 nm, and 340/460 nm for azimuthal, radial, and axial polarizations, respectively. This is due to the effective refractive index difference in three different polarizations. In  Fig. \ref{fig3} (a), the maximum $\eta$-value of  29\% is found for the radial polarization, which is in good agreement with the previously reported works in Refs. \cite{le2005spontaneous,klimov2004spontaneous}. The average $\eta$-value is 20\%, assuming the randomly oriented dipoles, which is in good agreement with the experimentally reported $\eta$-value in Ref. \cite{yalla2012efficient}. In  Fig. \ref{fig3} (b), the maximum $\eta$-value of  8\% is found for the radial polarization. The average $\eta$-value is 5\%, considering the randomly oriented dipoles for the experimental feasibility. The difference in the maximum $\eta$-value is due to the weak field confinement around the ONF. However, the ONF in water medium will find various applications for bio-sensing \cite{mauranyapin2017evanescent}.

As seen in Figs. \ref{fig4} (a) and (b), one can readily see the maximum and minimum $\eta$-values are for the radial and axial polarizations for both mediums, respectively. This is due to the axial component ($E_z$) of the electric field being low at the center of the NCF as discussed in Ref. \cite{le2004field}.  In Fig. \ref{fig4} (a), the maximum $\eta$-value for the radial polarization is almost the same for water and vacuum mediums up to $d^{in}$-value of 40 nm. However, a significant change for $\eta$-value occurred for the $d^{in}$-value from 50 nm to 300 nm for the vacuum compared to the water medium. This is due to the significant change in the effective refractive index of the NCF while the $d^{in}$-value is high.  In contrast, as seen in Fig. \ref{fig4} (b), one can readily see that as $d^{out}$-value increases the $\eta$-value for axial polarization increases. This may be due to the coupling to multi-modes (higher-order).

For the radial polarization, a maximum  $\eta$-value of 52\% is realized at  $d^{in}$-value of 100 nm and $d^{out}$-value of 360 nm of the NCF. In contrast, as seen in Fig. \ref{fig5} (a), one can readily see the maximum $\eta$-value for the radial and axial polarizations get affected when the $d^{in}$ is fixed at 250 nm. For the radial polarization, a maximum  $\eta$-value of 47\% is realized at the $d^{in}$-value of 250 nm and the $d^{out}$-value of 380 nm of the NCF. This is due to the weak field confinement inside the NCF because of the effective refractive index change as the inner diameter increases. We also simulated $\eta$-value for the azimuthal polarization and found the same result as the radial polarization. This is due to the placement of the SDS at the center in contrast to the SDS on the surface of the ONF. Note that $\eta$-value of 52\% is almost two times the values compared to the SDS placed on the surface of the ONF \cite{le2005spontaneous,klimov2004spontaneous}. The present value is almost 1.4 times the value compared to the SDS placed on the facet of the ONFT. Compared to other existing fiber-based platforms, the NCF is a promising avenue.  The average $\eta$-value is 35\%, assuming the randomly oriented dipoles, which is higher than the experimentally reported $\eta$-value for the ONF in Ref. \cite{yalla2012efficient}.

Two key challenges are to be achieved regarding the experimental feasibility of the present simulations. One is the fabrication of the NCF with such a small capillary hole and the other is placing a single quantum emitter inside the NCF. Commercial capillary optical fibers can be tapered to sub-wavelength diameters to realize NCF using a heat and pull technique \cite{ward2006heat,nayak2018nanofiber,tong2003subwavelength}. The placement of a single quantum emitter inside the NCF can be achieved by flowing through the center of the NCF by pushing the liquid from one end of the NCF. Considering the experimental realization of the simulation results, we choose water medium as quantum dots are dissolved in water. Regarding the sensitivity of the inner and outer diameters of the NCF to the $\eta$-value, one can readily see in Figs. \ref{fig4} and Fig. \ref{fig5} (a) that the peak $\eta$-values occurred in rather broad, not sharp. This implies that if there is any slight ambiguity in the inner and outer diameters of the NCF, it doesn't affect the $\eta$-value significantly. This would give additional freedom in the experimental realization.

As seen in Fig. \ref{fig5} (b), the $\eta$-value is almost unchanged while changing the $r^{in}$-value for the radial polarization. This suggests that the SDS position inside the NCF is not necessarily at the center to realize the maximum interaction i.e. maximum $\eta$-value. This would enable the easy experiments to place a single quantum emitter inside the NCF, as it is challenging to control the position of a single quantum emitter inside the NCF. Note that no significant change occurs for azimuthal polarization while changing the $r^{in}$-values up to 50 nm. In any case, the maximum $\eta$-value of more than 50\% can be realized using the NCF.

The $\eta$-value can be further enhanced in two possible ways: one is creating a cavity structure on the NCF \cite{yalla2014cavity,yalla2022one,yalla2020design,li2017optical}, and the other is increasing the refractive index of the medium and material. Regarding cavity, simulations suggest that $\eta$-value can be as high as 80\%, which is higher than the experimentally reported values of 65\% if the ONF is replaced with the NCF. Regarding increasing refractive index, simulations suggest that $\eta$-value can be higher than the present value if the silica is replaced with other high refractive index material like diamond/silicon nitride/gallium phosphate \cite{das2023efficient}. Also, the refractive index of the medium can affect the $\eta$-value if we replace water with any higher refractive index liquid \cite{faez2014coherent}.

\section{Conclusion}\label{sec5}
In summary, we reported numerically the efficient channeling of single photons from a single quantum emitter into guided modes of NCFs. The maximum channeling efficiency of $52$\% is found when the radially polarized dipole is placed at the center of the NCF in the liquid (water) medium. The optimum inner and outer diameters of the NCF are $100$ nm and $360$ nm, respectively for the emission wavelength of $620$ nm. Additionally, we investigated the SDS position dependence inside the NCF considering experimental discrepancies in placing the single quantum emitter inside the NCF. We found that the channeling efficiency remains almost the same for the water medium at the optimum condition. The present platform may open a novel route for quantum technologies and bio-sensing.

\backmatter

\bmhead{Acknowledgments}
RRY acknowledges financial support from the Scheme for Transformational and Advanced Research in Sciences (STARS) grant from the Indian Institute of Science (IISc), Ministry of Human Resource Development (MHRD) (File No. STARS/APR2019/PS/271/FS) and the Institute of Eminence (IoE) grant at the University of Hyderabad, Ministry of Education (MoE) (File No. RC2-21-019). RM acknowledges the University Grants Commission (UGC) for the financial support (Ref. No.:1412/CSIR-UGC NET June 2019).
\bmhead{Author contributions}
BE performed the numerical calculations and plotted the graphs. BE wrote the manuscript under the supervision of RRY. All authors reviewed the manuscript. This work forms a part of the PhD thesis of BE.

\section*{Declarations}

\bmhead{Ethical Approval}
Not applicable. No human and/or animal studies have been executed.

\bmhead{Funding}
To prepare this manuscript, funding was received by Institute of Eminence (IoE) grant at the University of Hyderabad, Ministry of Education (MoE) (File No. RC2-21-019).

\bmhead{Availability of data and materials}
Data underlying the results presented in this paper are not publicly available at this time but may be obtained from the authors upon reasonable request.

\bmhead{Confict of interest}
The authors declare no confict interests.

\bmhead{Competing interests}
The authors declare no competing interests.

\bibliographystyle{sn-mathphys.bst}
\bibliography{References}


\begin{thebibliography}{40}
\ifx \bisbn   \undefined \def \bisbn  #1{ISBN #1}\fi
\ifx \binits  \undefined \def \binits#1{#1}\fi
\ifx \bauthor  \undefined \def \bauthor#1{#1}\fi
\ifx \batitle  \undefined \def \batitle#1{#1}\fi
\ifx \bjtitle  \undefined \def \bjtitle#1{#1}\fi
\ifx \bvolume  \undefined \def \bvolume#1{\textbf{#1}}\fi
\ifx \byear  \undefined \def \byear#1{#1}\fi
\ifx \bissue  \undefined \def \bissue#1{#1}\fi
\ifx \bfpage  \undefined \def \bfpage#1{#1}\fi
\ifx \blpage  \undefined \def \blpage #1{#1}\fi
\ifx \burl  \undefined \def \burl#1{\textsf{#1}}\fi
\ifx \doiurl  \undefined \def \doiurl#1{\url{https://doi.org/#1}}\fi
\ifx \betal  \undefined \def \betal{\textit{et al.}}\fi
\ifx \binstitute  \undefined \def \binstitute#1{#1}\fi
\ifx \binstitutionaled  \undefined \def \binstitutionaled#1{#1}\fi
\ifx \bctitle  \undefined \def \bctitle#1{#1}\fi
\ifx \beditor  \undefined \def \beditor#1{#1}\fi
\ifx \bpublisher  \undefined \def \bpublisher#1{#1}\fi
\ifx \bbtitle  \undefined \def \bbtitle#1{#1}\fi
\ifx \bedition  \undefined \def \bedition#1{#1}\fi
\ifx \bseriesno  \undefined \def \bseriesno#1{#1}\fi
\ifx \blocation  \undefined \def \blocation#1{#1}\fi
\ifx \bsertitle  \undefined \def \bsertitle#1{#1}\fi
\ifx \bsnm \undefined \def \bsnm#1{#1}\fi
\ifx \bsuffix \undefined \def \bsuffix#1{#1}\fi
\ifx \bparticle \undefined \def \bparticle#1{#1}\fi
\ifx \barticle \undefined \def \barticle#1{#1}\fi
\bibcommenthead
\ifx \bconfdate \undefined \def \bconfdate #1{#1}\fi
\ifx \botherref \undefined \def \botherref #1{#1}\fi
\ifx \url \undefined \def \url#1{\textsf{#1}}\fi
\ifx \bchapter \undefined \def \bchapter#1{#1}\fi
\ifx \bbook \undefined \def \bbook#1{#1}\fi
\ifx \bcomment \undefined \def \bcomment#1{#1}\fi
\ifx \oauthor \undefined \def \oauthor#1{#1}\fi
\ifx \citeauthoryear \undefined \def \citeauthoryear#1{#1}\fi
\ifx \endbibitem  \undefined \def \endbibitem {}\fi
\ifx \bconflocation  \undefined \def \bconflocation#1{#1}\fi
\ifx \arxivurl  \undefined \def \arxivurl#1{\textsf{#1}}\fi
\csname PreBibitemsHook\endcsname

\bibitem[\protect\citeauthoryear{Hood et~al.}{2000}]{hood2000atom}
\begin{barticle}
\bauthor{\bsnm{Hood}, \binits{C.J.}},
\bauthor{\bsnm{Lynn}, \binits{T.}},
\bauthor{\bsnm{Doherty}, \binits{A.}},
\bauthor{\bsnm{Parkins}, \binits{A.}},
\bauthor{\bsnm{Kimble}, \binits{H.}}:
\batitle{The atom-cavity microscope: Single atoms bound in orbit by single
  photons}.
\bjtitle{Science}
\bvolume{287}(\bissue{5457}),
\bfpage{1447}--\blpage{1453}
(\byear{2000})
\end{barticle}
\endbibitem

\bibitem[\protect\citeauthoryear{Barros et~al.}{2009}]{barros2009deterministic}
\begin{barticle}
\bauthor{\bsnm{Barros}, \binits{H.}},
\bauthor{\bsnm{Stute}, \binits{A.}},
\bauthor{\bsnm{Northup}, \binits{T.}},
\bauthor{\bsnm{Russo}, \binits{C.}},
\bauthor{\bsnm{Schmidt}, \binits{P.}},
\bauthor{\bsnm{Blatt}, \binits{R.}}:
\batitle{Deterministic single-photon source from a single ion}.
\bjtitle{New Journal of Physics}
\bvolume{11}(\bissue{10}),
\bfpage{103004}
(\byear{2009})
\end{barticle}
\endbibitem

\bibitem[\protect\citeauthoryear{Fleury et~al.}{2000}]{fleury2000nonclassical}
\begin{barticle}
\bauthor{\bsnm{Fleury}, \binits{L.}},
\bauthor{\bsnm{Segura}, \binits{J.-M.}},
\bauthor{\bsnm{Zumofen}, \binits{G.}},
\bauthor{\bsnm{Hecht}, \binits{B.}},
\bauthor{\bsnm{Wild}, \binits{U.}}:
\batitle{Nonclassical photon statistics in single-molecule fluorescence at room
  temperature}.
\bjtitle{Physical review letters}
\bvolume{84}(\bissue{6}),
\bfpage{1148}
(\byear{2000})
\end{barticle}
\endbibitem

\bibitem[\protect\citeauthoryear{Michler et~al.}{2000}]{michler2000quantum}
\begin{barticle}
\bauthor{\bsnm{Michler}, \binits{P.}},
\bauthor{\bsnm{Imamo{\u{g}}lu}, \binits{A.}},
\bauthor{\bsnm{Mason}, \binits{M.}},
\bauthor{\bsnm{Carson}, \binits{P.}},
\bauthor{\bsnm{Strouse}, \binits{G.}},
\bauthor{\bsnm{Buratto}, \binits{S.}}:
\batitle{Quantum correlation among photons from a single quantum dot at room
  temperature}.
\bjtitle{Nature}
\bvolume{406}(\bissue{6799}),
\bfpage{968}--\blpage{970}
(\byear{2000})
\end{barticle}
\endbibitem

\bibitem[\protect\citeauthoryear{Kurtsiefer
  et~al.}{2000}]{kurtsiefer2000stable}
\begin{barticle}
\bauthor{\bsnm{Kurtsiefer}, \binits{C.}},
\bauthor{\bsnm{Mayer}, \binits{S.}},
\bauthor{\bsnm{Zarda}, \binits{P.}},
\bauthor{\bsnm{Weinfurter}, \binits{H.}}:
\batitle{Stable solid-state source of single photons}.
\bjtitle{Physical review letters}
\bvolume{85}(\bissue{2}),
\bfpage{290}
(\byear{2000})
\end{barticle}
\endbibitem

\bibitem[\protect\citeauthoryear{Aharonovich
  et~al.}{2016}]{aharonovich2016solid}
\begin{barticle}
\bauthor{\bsnm{Aharonovich}, \binits{I.}},
\bauthor{\bsnm{Englund}, \binits{D.}},
\bauthor{\bsnm{Toth}, \binits{M.}}:
\batitle{Solid-state single-photon emitters}.
\bjtitle{Nature photonics}
\bvolume{10}(\bissue{10}),
\bfpage{631}--\blpage{641}
(\byear{2016})
\end{barticle}
\endbibitem

\bibitem[\protect\citeauthoryear{Kimble}{2008}]{kimble2008quantum}
\begin{barticle}
\bauthor{\bsnm{Kimble}, \binits{H.J.}}:
\batitle{The quantum internet}.
\bjtitle{Nature}
\bvolume{453}(\bissue{7198}),
\bfpage{1023}--\blpage{1030}
(\byear{2008})
\end{barticle}
\endbibitem

\bibitem[\protect\citeauthoryear{Bremer et~al.}{2022}]{bremer2022fiber}
\begin{botherref}
\oauthor{\bsnm{Bremer}, \binits{L.}},
\oauthor{\bsnm{Rodt}, \binits{S.}},
\oauthor{\bsnm{Reitzenstein}, \binits{S.}}:
Fiber-coupled quantum light sources based on solid-state quantum emitters.
Materials for Quantum Technology
(2022)
\end{botherref}
\endbibitem

\bibitem[\protect\citeauthoryear{Solomon et~al.}{2001}]{solomon2001single}
\begin{barticle}
\bauthor{\bsnm{Solomon}, \binits{G.}},
\bauthor{\bsnm{Pelton}, \binits{M.}},
\bauthor{\bsnm{Yamamoto}, \binits{Y.}}:
\batitle{Single-mode spontaneous emission from a single quantum dot in a
  three-dimensional microcavity}.
\bjtitle{Physical Review Letters}
\bvolume{86}(\bissue{17}),
\bfpage{3903}
(\byear{2001})
\end{barticle}
\endbibitem

\bibitem[\protect\citeauthoryear{Schr{\"o}der
  et~al.}{2011}]{schroder2011ultrabright}
\begin{barticle}
\bauthor{\bsnm{Schr{\"o}der}, \binits{T.}},
\bauthor{\bsnm{G{\"a}deke}, \binits{F.}},
\bauthor{\bsnm{Banholzer}, \binits{M.J.}},
\bauthor{\bsnm{Benson}, \binits{O.}}:
\batitle{Ultrabright and efficient single-photon generation based on
  nitrogen-vacancy centres in nanodiamonds on a solid immersion lens}.
\bjtitle{New Journal of Physics}
\bvolume{13}(\bissue{5}),
\bfpage{055017}
(\byear{2011})
\end{barticle}
\endbibitem

\bibitem[\protect\citeauthoryear{Shambat et~al.}{2011}]{shambat2011optical}
\begin{barticle}
\bauthor{\bsnm{Shambat}, \binits{G.}},
\bauthor{\bsnm{Provine}, \binits{J.}},
\bauthor{\bsnm{Rivoire}, \binits{K.}},
\bauthor{\bsnm{Sarmiento}, \binits{T.}},
\bauthor{\bsnm{Harris}, \binits{J.}},
\bauthor{\bsnm{Vu{\v{c}}kovi{\'c}}, \binits{J.}}:
\batitle{Optical fiber tips functionalized with semiconductor photonic crystal
  cavities}.
\bjtitle{Applied Physics Letters}
\bvolume{99}(\bissue{19}),
\bfpage{191102}
(\byear{2011})
\end{barticle}
\endbibitem

\bibitem[\protect\citeauthoryear{Patel et~al.}{2016}]{patel2016efficient}
\begin{barticle}
\bauthor{\bsnm{Patel}, \binits{R.N.}},
\bauthor{\bsnm{Schr{\"o}der}, \binits{T.}},
\bauthor{\bsnm{Wan}, \binits{N.}},
\bauthor{\bsnm{Li}, \binits{L.}},
\bauthor{\bsnm{Mouradian}, \binits{S.L.}},
\bauthor{\bsnm{Chen}, \binits{E.H.}},
\bauthor{\bsnm{Englund}, \binits{D.R.}}:
\batitle{Efficient photon coupling from a diamond nitrogen vacancy center by
  integration with silica fiber}.
\bjtitle{Light: Science \& Applications}
\bvolume{5}(\bissue{2}),
\bfpage{16032}--\blpage{16032}
(\byear{2016})
\end{barticle}
\endbibitem

\bibitem[\protect\citeauthoryear{Akimov et~al.}{2007}]{akimov2007generation}
\begin{barticle}
\bauthor{\bsnm{Akimov}, \binits{A.}},
\bauthor{\bsnm{Mukherjee}, \binits{A.}},
\bauthor{\bsnm{Yu}, \binits{C.}},
\bauthor{\bsnm{Chang}, \binits{D.}},
\bauthor{\bsnm{Zibrov}, \binits{A.}},
\bauthor{\bsnm{Hemmer}, \binits{P.}},
\bauthor{\bsnm{Park}, \binits{H.}},
\bauthor{\bsnm{Lukin}, \binits{M.}}:
\batitle{Generation of single optical plasmons in metallic nanowires coupled to
  quantum dots}.
\bjtitle{Nature}
\bvolume{450}(\bissue{7168}),
\bfpage{402}--\blpage{406}
(\byear{2007})
\end{barticle}
\endbibitem

\bibitem[\protect\citeauthoryear{Nayak et~al.}{2018}]{nayak2018nanofiber}
\begin{barticle}
\bauthor{\bsnm{Nayak}, \binits{K.P.}},
\bauthor{\bsnm{Sadgrove}, \binits{M.}},
\bauthor{\bsnm{Yalla}, \binits{R.}},
\bauthor{\bsnm{Le~Kien}, \binits{F.}},
\bauthor{\bsnm{Hakuta}, \binits{K.}}:
\batitle{Nanofiber quantum photonics}.
\bjtitle{Journal of Optics}
\bvolume{20}(\bissue{7}),
\bfpage{073001}
(\byear{2018})
\end{barticle}
\endbibitem

\bibitem[\protect\citeauthoryear{Vetsch et~al.}{2010}]{vetsch2010optical}
\begin{barticle}
\bauthor{\bsnm{Vetsch}, \binits{E.}},
\bauthor{\bsnm{Reitz}, \binits{D.}},
\bauthor{\bsnm{Sagu{\'e}}, \binits{G.}},
\bauthor{\bsnm{Schmidt}, \binits{R.}},
\bauthor{\bsnm{Dawkins}, \binits{S.}},
\bauthor{\bsnm{Rauschenbeutel}, \binits{A.}}:
\batitle{Optical interface created by laser-cooled atoms trapped in the
  evanescent field surrounding an optical nanofiber}.
\bjtitle{Physical review letters}
\bvolume{104}(\bissue{20}),
\bfpage{203603}
(\byear{2010})
\end{barticle}
\endbibitem

\bibitem[\protect\citeauthoryear{Yalla et~al.}{2012}]{yalla2012efficient}
\begin{barticle}
\bauthor{\bsnm{Yalla}, \binits{R.}},
\bauthor{\bsnm{Le~Kien}, \binits{F.}},
\bauthor{\bsnm{Morinaga}, \binits{M.}},
\bauthor{\bsnm{Hakuta}, \binits{K.}}:
\batitle{Efficient channeling of fluorescence photons from single quantum dots
  into guided modes of optical nanofiber}.
\bjtitle{Physical review letters}
\bvolume{109}(\bissue{6}),
\bfpage{063602}
(\byear{2012})
\end{barticle}
\endbibitem

\bibitem[\protect\citeauthoryear{Fujiwara et~al.}{2011}]{fujiwara2011highly}
\begin{barticle}
\bauthor{\bsnm{Fujiwara}, \binits{M.}},
\bauthor{\bsnm{Toubaru}, \binits{K.}},
\bauthor{\bsnm{Noda}, \binits{T.}},
\bauthor{\bsnm{Zhao}, \binits{H.-Q.}},
\bauthor{\bsnm{Takeuchi}, \binits{S.}}:
\batitle{Highly efficient coupling of photons from nanoemitters into
  single-mode optical fibers}.
\bjtitle{Nano letters}
\bvolume{11}(\bissue{10}),
\bfpage{4362}--\blpage{4365}
(\byear{2011})
\end{barticle}
\endbibitem

\bibitem[\protect\citeauthoryear{Le~Kien et~al.}{2005}]{le2005spontaneous}
\begin{barticle}
\bauthor{\bsnm{Le~Kien}, \binits{F.}},
\bauthor{\bsnm{Gupta}, \binits{S.D.}},
\bauthor{\bsnm{Balykin}, \binits{V.}},
\bauthor{\bsnm{Hakuta}, \binits{K.}}:
\batitle{Spontaneous emission of a cesium atom near a nanofiber: Efficient
  coupling of light to guided modes}.
\bjtitle{Physical Review A}
\bvolume{72}(\bissue{3}),
\bfpage{032509}
(\byear{2005})
\end{barticle}
\endbibitem

\bibitem[\protect\citeauthoryear{Klimov and
  Ducloy}{2004}]{klimov2004spontaneous}
\begin{barticle}
\bauthor{\bsnm{Klimov}, \binits{V.V.}},
\bauthor{\bsnm{Ducloy}, \binits{M.}}:
\batitle{Spontaneous emission rate of an excited atom placed near a nanofiber}.
\bjtitle{Physical Review A}
\bvolume{69}(\bissue{1}),
\bfpage{013812}
(\byear{2004})
\end{barticle}
\endbibitem

\bibitem[\protect\citeauthoryear{Chonan et~al.}{2014}]{chonan2014efficient}
\begin{barticle}
\bauthor{\bsnm{Chonan}, \binits{S.}},
\bauthor{\bsnm{Kato}, \binits{S.}},
\bauthor{\bsnm{Aoki}, \binits{T.}}:
\batitle{Efficient single-mode photon-coupling device utilizing a nanofiber
  tip}.
\bjtitle{Scientific reports}
\bvolume{4}(\bissue{1}),
\bfpage{4785}
(\byear{2014})
\end{barticle}
\endbibitem

\bibitem[\protect\citeauthoryear{Yonezu et~al.}{2017}]{yonezu2017efficient}
\begin{barticle}
\bauthor{\bsnm{Yonezu}, \binits{Y.}},
\bauthor{\bsnm{Wakui}, \binits{K.}},
\bauthor{\bsnm{Furusawa}, \binits{K.}},
\bauthor{\bsnm{Takeoka}, \binits{M.}},
\bauthor{\bsnm{Semba}, \binits{K.}},
\bauthor{\bsnm{Aoki}, \binits{T.}}:
\batitle{Efficient single-photon coupling from a nitrogen-vacancy center
  embedded in a diamond nanowire utilizing an optical nanofiber}.
\bjtitle{Scientific reports}
\bvolume{7}(\bissue{1}),
\bfpage{12985}
(\byear{2017})
\end{barticle}
\endbibitem

\bibitem[\protect\citeauthoryear{Morrissey
  et~al.}{2013}]{morrissey2013spectroscopy}
\begin{barticle}
\bauthor{\bsnm{Morrissey}, \binits{M.J.}},
\bauthor{\bsnm{Deasy}, \binits{K.}},
\bauthor{\bsnm{Frawley}, \binits{M.}},
\bauthor{\bsnm{Kumar}, \binits{R.}},
\bauthor{\bsnm{Prel}, \binits{E.}},
\bauthor{\bsnm{Russell}, \binits{L.}},
\bauthor{\bsnm{Truong}, \binits{V.G.}},
\bauthor{\bsnm{Chormaic}, \binits{S.N.}}:
\batitle{Spectroscopy, manipulation and trapping of neutral atoms, molecules,
  and other particles using optical nanofibers: a review}.
\bjtitle{Sensors}
\bvolume{13}(\bissue{8}),
\bfpage{10449}--\blpage{10481}
(\byear{2013})
\end{barticle}
\endbibitem

\bibitem[\protect\citeauthoryear{Yang et~al.}{2023}]{yang2023generating}
\begin{barticle}
\bauthor{\bsnm{Yang}, \binits{L.}},
\bauthor{\bsnm{Zhou}, \binits{Z.}},
\bauthor{\bsnm{Wu}, \binits{H.}},
\bauthor{\bsnm{Dang}, \binits{H.}},
\bauthor{\bsnm{Yang}, \binits{Y.}},
\bauthor{\bsnm{Gao}, \binits{J.}},
\bauthor{\bsnm{Guo}, \binits{X.}},
\bauthor{\bsnm{Wang}, \binits{P.}},
\bauthor{\bsnm{Tong}, \binits{L.}}:
\batitle{Generating a sub-nanometer-confined optical field in a nanoslit
  waveguiding mode}.
\bjtitle{Advanced Photonics}
\bvolume{5}(\bissue{4}),
\bfpage{046003}--\blpage{046003}
(\byear{2023})
\end{barticle}
\endbibitem

\bibitem[\protect\citeauthoryear{Yalla et~al.}{2022}]{yalla2022integration}
\begin{botherref}
\oauthor{\bsnm{Yalla}, \binits{R.}},
\oauthor{\bsnm{Kojima}, \binits{Y.}},
\oauthor{\bsnm{Fukumoto}, \binits{Y.}},
\oauthor{\bsnm{Suzuki}, \binits{H.}},
\oauthor{\bsnm{Ariyada}, \binits{O.}},
\oauthor{\bsnm{Shafi}, \binits{K.M.}},
\oauthor{\bsnm{Nayak}, \binits{K.P.}},
\oauthor{\bsnm{Hakuta}, \binits{K.}}:
Integration of silicon-vacancy centers in nanodiamonds with an optical
  nanofiber.
Applied Physics Letters
\textbf{120}(24)
(2022)
\end{botherref}
\endbibitem

\bibitem[\protect\citeauthoryear{Resmi et~al.}{2023}]{resmi2023efficient}
\begin{bchapter}
\bauthor{\bsnm{Resmi}, \binits{M.}},
\bauthor{\bsnm{Bashaiah}, \binits{E.}},
\bauthor{\bsnm{Das}, \binits{B.}},
\bauthor{\bsnm{Yalla}, \binits{R.}}:
\bctitle{Efficient fiber-coupled single photon source using an optical
  nanofiber tip}.
In: \bbtitle{Women in Optics and Photonics in India 2022},
vol. \bseriesno{12638},
pp. \bfpage{107}--\blpage{109}
(\byear{2023}).
\bcomment{SPIE}
\end{bchapter}
\endbibitem

\bibitem[\protect\citeauthoryear{Das et~al.}{2023}]{das2023efficient}
\begin{bchapter}
\bauthor{\bsnm{Das}, \binits{B.}},
\bauthor{\bsnm{Resmi}, \binits{M.}},
\bauthor{\bsnm{Bashaiah}, \binits{E.}},
\bauthor{\bsnm{Yalla}, \binits{R.}}:
\bctitle{Efficient single-mode coupling design using a silica/diamond nano-tip
  with a gold nanoparticle}.
In: \bbtitle{Women in Optics and Photonics in India 2022},
vol. \bseriesno{12638},
pp. \bfpage{125}--\blpage{127}
(\byear{2023}).
\bcomment{SPIE}
\end{bchapter}
\endbibitem

\bibitem[\protect\citeauthoryear{Resmi et~al.}{2024}]{m2024channeling}
\begin{botherref}
\oauthor{\bsnm{Resmi}, \binits{M.}},
\oauthor{\bsnm{Elaganuru}, \binits{B.}},
\oauthor{\bsnm{Ramachandrarao}, \binits{Y.}}:
Channeling of fluorescence photons from quantum dots into guided modes of an
  optical nanofiber tip.
arXiv
\textbf{2401.16891}
(2024)
\end{botherref}
\endbibitem

\bibitem[\protect\citeauthoryear{Yalla et~al.}{2014}]{yalla2014cavity}
\begin{barticle}
\bauthor{\bsnm{Yalla}, \binits{R.}},
\bauthor{\bsnm{Sadgrove}, \binits{M.}},
\bauthor{\bsnm{Nayak}, \binits{K.P.}},
\bauthor{\bsnm{Hakuta}, \binits{K.}}:
\batitle{Cavity quantum electrodynamics on a nanofiber using a composite
  photonic crystal cavity}.
\bjtitle{Physical review letters}
\bvolume{113}(\bissue{14}),
\bfpage{143601}
(\byear{2014})
\end{barticle}
\endbibitem

\bibitem[\protect\citeauthoryear{Faez et~al.}{2014}]{faez2014coherent}
\begin{barticle}
\bauthor{\bsnm{Faez}, \binits{S.}},
\bauthor{\bsnm{T{\"u}rschmann}, \binits{P.}},
\bauthor{\bsnm{Haakh}, \binits{H.R.}},
\bauthor{\bsnm{G{\"o}tzinger}, \binits{S.}},
\bauthor{\bsnm{Sandoghdar}, \binits{V.}}:
\batitle{Coherent interaction of light and single molecules in a dielectric
  nanoguide}.
\bjtitle{Physical review letters}
\bvolume{113}(\bissue{21}),
\bfpage{213601}
(\byear{2014})
\end{barticle}
\endbibitem

\bibitem[\protect\citeauthoryear{Taflove
  et~al.}{2005}]{taflove2005computational}
\begin{barticle}
\bauthor{\bsnm{Taflove}, \binits{A.}},
\bauthor{\bsnm{Hagness}, \binits{S.C.}},
\bauthor{\bsnm{Piket-May}, \binits{M.}}:
\batitle{Computational electromagnetics: the finite-difference time-domain
  method}.
\bjtitle{The Electrical Engineering Handbook}
\bvolume{3}(\bissue{629-670}),
\bfpage{15}
(\byear{2005})
\end{barticle}
\endbibitem

\bibitem[\protect\citeauthoryear{Schneider}{2010}]{schneider2010understanding}
\begin{botherref}
\oauthor{\bsnm{Schneider}, \binits{J.B.}}:
Understanding the finite-difference time-domain method.
School of electrical engineering and computer science Washington State
  University
\textbf{28}
(2010)
\end{botherref}
\endbibitem

\bibitem[\protect\citeauthoryear{Shafi et~al.}{2018}]{shafi2018hybrid}
\begin{barticle}
\bauthor{\bsnm{Shafi}, \binits{K.M.}},
\bauthor{\bsnm{Luo}, \binits{W.}},
\bauthor{\bsnm{Yalla}, \binits{R.}},
\bauthor{\bsnm{Iida}, \binits{K.}},
\bauthor{\bsnm{Tsutsumi}, \binits{E.}},
\bauthor{\bsnm{Miyanaga}, \binits{A.}},
\bauthor{\bsnm{Hakuta}, \binits{K.}}:
\batitle{Hybrid system of an optical nanofibre and a single quantum dot
  operated at cryogenic temperatures}.
\bjtitle{Scientific reports}
\bvolume{8}(\bissue{1}),
\bfpage{13494}
(\byear{2018})
\end{barticle}
\endbibitem

\bibitem[\protect\citeauthoryear{Le~Kien et~al.}{2004}]{le2004field}
\begin{barticle}
\bauthor{\bsnm{Le~Kien}, \binits{F.}},
\bauthor{\bsnm{Liang}, \binits{J.}},
\bauthor{\bsnm{Hakuta}, \binits{K.}},
\bauthor{\bsnm{Balykin}, \binits{V.}}:
\batitle{Field intensity distributions and polarization orientations in a
  vacuum-clad subwavelength-diameter optical fiber}.
\bjtitle{Optics Communications}
\bvolume{242}(\bissue{4-6}),
\bfpage{445}--\blpage{455}
(\byear{2004})
\end{barticle}
\endbibitem

\bibitem[\protect\citeauthoryear{Okamoto}{2021}]{okamoto2021fundamentals}
\begin{bbook}
\bauthor{\bsnm{Okamoto}, \binits{K.}}:
\bbtitle{Fundamentals of Optical Waveguides}.
\bpublisher{Elsevier}, \blocation{???}
(\byear{2021})
\end{bbook}
\endbibitem

\bibitem[\protect\citeauthoryear{Mauranyapin
  et~al.}{2017}]{mauranyapin2017evanescent}
\begin{barticle}
\bauthor{\bsnm{Mauranyapin}, \binits{N.}},
\bauthor{\bsnm{Madsen}, \binits{L.}},
\bauthor{\bsnm{Taylor}, \binits{M.}},
\bauthor{\bsnm{Waleed}, \binits{M.}},
\bauthor{\bsnm{Bowen}, \binits{W.}}:
\batitle{Evanescent single-molecule biosensing with quantum-limited precision}.
\bjtitle{Nature Photonics}
\bvolume{11}(\bissue{8}),
\bfpage{477}--\blpage{481}
(\byear{2017})
\end{barticle}
\endbibitem

\bibitem[\protect\citeauthoryear{Ward et~al.}{2006}]{ward2006heat}
\begin{botherref}
\oauthor{\bsnm{Ward}, \binits{J.M.}},
\oauthor{\bsnm{O’Shea}, \binits{D.G.}},
\oauthor{\bsnm{Shortt}, \binits{B.J.}},
\oauthor{\bsnm{Morrissey}, \binits{M.J.}},
\oauthor{\bsnm{Deasy}, \binits{K.}},
\oauthor{\bsnm{Nic~Chormaic}, \binits{S.G.}}:
Heat-and-pull rig for fiber taper fabrication.
Review of scientific instruments
\textbf{77}(8)
(2006)
\end{botherref}
\endbibitem

\bibitem[\protect\citeauthoryear{Tong et~al.}{2003}]{tong2003subwavelength}
\begin{barticle}
\bauthor{\bsnm{Tong}, \binits{L.}},
\bauthor{\bsnm{Gattass}, \binits{R.R.}},
\bauthor{\bsnm{Ashcom}, \binits{J.B.}},
\bauthor{\bsnm{He}, \binits{S.}},
\bauthor{\bsnm{Lou}, \binits{J.}},
\bauthor{\bsnm{Shen}, \binits{M.}},
\bauthor{\bsnm{Maxwell}, \binits{I.}},
\bauthor{\bsnm{Mazur}, \binits{E.}}:
\batitle{Subwavelength-diameter silica wires for low-loss optical wave
  guiding}.
\bjtitle{Nature}
\bvolume{426}(\bissue{6968}),
\bfpage{816}--\blpage{819}
(\byear{2003})
\end{barticle}
\endbibitem

\bibitem[\protect\citeauthoryear{Yalla et~al.}{2022}]{yalla2022one}
\begin{botherref}
\oauthor{\bsnm{Yalla}, \binits{R.}},
\oauthor{\bsnm{Muhammed~Shafi}, \binits{K.}},
\oauthor{\bsnm{Nayak}, \binits{K.P.}},
\oauthor{\bsnm{Hakuta}, \binits{K.}}:
One-sided composite cavity on an optical nanofiber for cavity qed.
Applied Physics Letters
\textbf{120}(7)
(2022)
\end{botherref}
\endbibitem

\bibitem[\protect\citeauthoryear{Yalla and Hakuta}{2020}]{yalla2020design}
\begin{barticle}
\bauthor{\bsnm{Yalla}, \binits{R.}},
\bauthor{\bsnm{Hakuta}, \binits{K.}}:
\batitle{Design and implementation of a tunable composite photonic crystal
  cavity on an optical nanofiber}.
\bjtitle{Applied Physics B}
\bvolume{126}(\bissue{11}),
\bfpage{187}
(\byear{2020})
\end{barticle}
\endbibitem

\bibitem[\protect\citeauthoryear{Li et~al.}{2017}]{li2017optical}
\begin{botherref}
\oauthor{\bsnm{Li}, \binits{W.}},
\oauthor{\bsnm{Du}, \binits{J.}},
\oauthor{\bsnm{Truong}, \binits{V.G.}},
\oauthor{\bsnm{Nic~Chormaic}, \binits{S.}}:
Optical nanofiber-based cavity induced by periodic air-nanohole arrays.
Applied Physics Letters
\textbf{110}(25)
(2017)
\end{botherref}
\endbibitem

\end{thebibliography}

\end{document}